\documentclass[fleqn,twoside]{article}
\usepackage{espcrc2}

\usepackage{graphicx}
\usepackage[figuresright]{rotating}

\newcommand{\AmS}{{\protect\the\textfont2
  A\kern-.1667em\lower.5ex\hbox{M}\kern-.125emS}}

\title{BooNE Has Begun}

\author{E. D. Zimmerman\address{University of Colorado\\ Boulder, Colorado 80309}%
        \thanks{For the BooNE Collaboration.}}
       
\begin{document}

\begin{abstract}
E898, the MiniBooNE experiment at Fermi National Accelerator Laboratory, has 
begun data collection. The experiment will test the neutrino oscillation 
signal reported by the Liquid Scintillator Neutrino Detector at Los Alamos
National Laboratory. Data collection began in late August 2002.
\vspace{1pc}
\end{abstract}

\maketitle

\section{PHYSICS INTRODUCTION: LSND and KARMEN}

There are now several sources of evidence for neutrino oscillations. 
There is evidence from solar neutrino experiments,
which observe fewer electron neutrinos than would be consistent with
the sun's energy output \cite{solar}. In addition, the recent SNO
results indicate that there is a non-electron flavor active neutrino
component in the solar flux \cite{sno,newsno}.  Evidence also comes from
experiments studying neutrinos produced by cosmic rays in the earth's
atmosphere, which find that muon neutrinos entering the detector from
below (which had to travel through the earth) are depleted when
compared with those incident from above (which traveled a shorter
distance) \cite{atmos}.  Finally, there is evidence from a lone
accelerator experiment, LSND at Los Alamos, which observed an excess
of $\bar\nu_e$ events from a predominantly $\bar\nu_\mu$ source
\cite{lsnd}. 

LSND used a beam-stop neutrino source at the 800~MeV LAMPF proton
accelerator. The primary source of neutrinos was $\pi^+$ and $\mu^+$
decays at rest (DAR) in the target, which yielded $\nu_\mu$, $\bar
\nu_\mu$, and $\nu_e$ with energies below 53~MeV. In addition, $\pi^+$
and $\pi^-$ decays in flight (DIF) provided a small flux of
higher-energy $\nu_\mu$ and $\bar\nu_\mu$. The $\bar\nu_e$ flux was
below $10^{-3}$ of the total DAR rate.  The LSND data were collected
between 1993 and 1998. The first data set, collected 1993-1995, used a
water target and provided of 59\% of the DAR flux; the remainder of
the running used a heavy metal target composed mostly of tungsten.
The collaboration searched for $\bar\nu_e$ appearance using the
reaction $\bar\nu_e p \rightarrow e^+n$ in a 167-ton
scintillator-doped mineral oil (CH$_2$) target/detector. The detector
sat 30~m from the target, providing an oscillation scale $L/E \sim
0.6-1$~m/MeV. The detector, which was instrumented with 1220 8-inch
photomultiplier tubes (PMTs), observed a \v{C}erenkov ring and
scintillation light from the positron emitted in the neutrino
interaction. An additional handle was the detection of the $2.2$~MeV
neutron-capture gamma ray from the reaction $np \rightarrow
d\gamma$. The appropriate delayed coincidence (the neutron capture
lifetime in oil is 186~$\mu$~s) and spatial correlation between the
$e^+$ and $\gamma$ were studied for DAR $\bar\nu_e$ candidates.

\begin{figure}[t]
\includegraphics[width=3in]{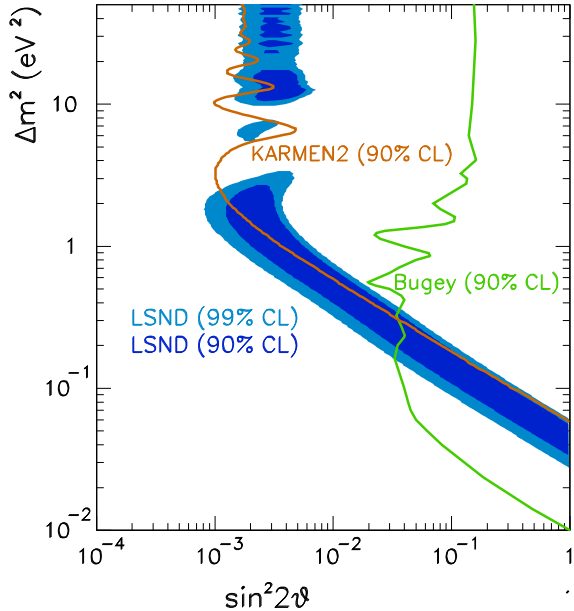}
\caption{LSND 90\% and 99\% confidence level allowed regions 
(shaded) along with KARMEN2 and Bugey 90\% confidence level exclusion
limits and expected MiniBooNE sensitivity.}
\label{lsndkar}
\end{figure}

In 2001, LSND presented the final oscillation search results, which
gave a total $\bar\nu_e$ excess above background of $87.9 \pm 22.4 \pm
6.0$ events in the DAR energy range. The dominant background was
beam-unrelated events, primarily from cosmic rays. These backgrounds
were measured using the 94\% of detector livetime when the beam was
not on. No significant signal was observed in DIF events; the total
$\nu_e$/$\bar\nu_e$ excess above background was $8.1 \pm 12.2 \pm 1.7$
events, consistent with the DAR result. The total events and energy
distributions of the DAR and DIF events were used to constrain the
oscillation parameter space (Fig.~\ref{lsndkar}). 

Another experiment of similar design, the Karlsruhe-Rutherford Medium
Energy Neutrino (KARMEN) experiment at the ISIS facility of the
Rutherford Laboratory, also searched for $\bar\nu_\mu \rightarrow
\bar\nu_e$ oscillations. KARMEN used a similar beam-stop neutrino
source, but with a segmented smaller neutrino target (56
tons). KARMEN's sensitivity was enhanced because the lower beam duty
factor ($10^{-5}$) allowed beam-unrelated events to be removed more
effectively with a timing cut. In addition, KARMEN had higher flux
because it was closer to the target (18~m versus 30~m).  This did,
however, reduce KARMEN's sensitivity to low-$\Delta m^2$ oscillations
compared to LSND. KARMEN's most recent published result \cite{karmen},
using data collected from 1997 to 2001, reported 15 $\bar\nu_e$ oscillation
candidates with an expected background of $15.8 \pm 0.5$ events. This
result does not provide evidence for oscillations, and indeed can be
used to rule out most of the high-$\Delta m^2$ portions of the LSND
allowed region. However, an analysis of the combined LSND and KARMEN
data sets has found regions of oscillation parameter space which 
fit both experiments' data well \cite{eitel}.

The LSND data indicate a much larger $\Delta m^2$ than atmospheric or
solar experiments: $\Delta m^2 \sim 0.1-10$~eV$^2$. This led to the
paradox of three $\Delta m^2$ values all of different orders of
magnitude; this is impossible if there are only three neutrino
masses. The more common way to account for all the existing
oscillation data is to introduce one or more ``sterile'' neutrino
flavors \cite{sterile}. A more recent idea, motivated by
extra-dimensions models, has been to introduce maximal $CPT$ violation
in the neutrino mass matrix, thereby giving neutrinos and
antineutrinos differing mass hierarchies \cite{cptbrane}.

\section{MiniBooNE OVERVIEW}

MiniBooNE (Experiment 898 at Fermilab) \cite{boone} is a
short-baseline neutrino oscillation experiment which is designed to
confirm or rule out LSND unequivocally.  It uses an 8~GeV proton beam
from the Fermilab Booster to produce pions, which are focused by a
horn into a decay pipe, where they decay in flight to produce a nearly
pure $\nu_\mu$ beam. The neutrinos are detected at a mineral oil
\v{C}erenkov detector 500~m away. The detector will use \v{C}erenkov
ring shape information to distinguish charged-current $\nu_\mu$ from
$\nu_e$ interactions, searching for an excess of $\nu_e$ which would
indicate oscillations. Data collection began in late summer 2002 and
is expected to last two to three years.  MiniBooNE is the first stage
of the BooNE program, which will continue with a two-detector
experiment to make precise measurements of oscillation parameters if
LSND is confirmed.

There are several major differences between MiniBooNE and LSND, which
should assure that systematic errors are independent. First, MiniBooNE
operates at an energy and oscillation baseline over an order of
magnitude greater than LSND: $E_\nu \sim 500-1000$~MeV, compared to
$30-53$~MeV at LSND. The baseline $L=500$~m, versus 30~m at
LSND. $L/E$ remains similar, ensuring that the oscillation sensitivity
is maximized in the same region of parameter space as LSND. MiniBooNE
uses the quasielestic neutrino scattering reaction 
$\nu_e {\rm ^{12}C} \rightarrow e^-X$ with the leading lepton's 
\v{C}erenkov ring reconstructed, rather than LSND's antineutrino
interaction with a hydrogen nucleus followed by neutron capture. Finally,
MiniBooNE's goal is a factor of ten higher statistics than LSND had. 

\section{BEAM DETAILS}

The BooNE neutrino beam begins with an 8~GeV primary proton beam from the
Fermilab Booster accelerator. The beam arrives in 1.6~$\mu$s
pulses, with five pulses per second. Within each pulse, the beam arrives
in 80 bunches (``RF buckets'') 19~ns apart. Protons from the primary beam strike a 71~cm beryllium target, producing
short lived hadrons with a typical transverse momentum of 0.3~GeV/$c$.
The hadrons are focused by the magnetic fields generated from
a high-current-carrying device called a ``horn.'' The target is
located within the magnetic focusing horn.  A horn was chosen because
it gives higher angular and momentum acceptances than other focusing
systems. It can be made to withstand high radiation levels, has
cylindrical symmetry, and also gives sign selection.

\begin{figure}[t]
\includegraphics[width=3in]{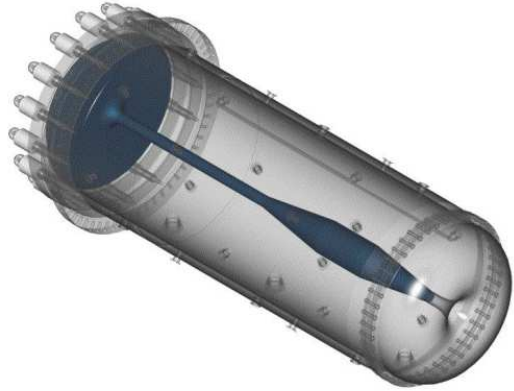}
\caption{A view of the focusing horn. Beam enters from the upper left.
All elements shown are aluminum.  The outer conductor is rendered transparent.
The target resides inside
the narrow neck of the upstream end of the inner conductor. The inner
conductor is six feet long, and the horn's diameter is two feet. {\em Image
by Bartoszek Engineering.}}
\label{horn}
\end{figure}

A horn (see Fig.~\ref{horn}) contains a pulsed toroidal magnetic field
in the volume between two coaxial conductors.  Current flows along the
inner (small radius) conductor and back along the outer (large radius)
conductor. There is no field inside the inner conductor, nor outside
the outer conductor.  In the volume between the inner and outer
conductors, the magnitude of the field is given by $B({\rm kG}) \approx
0.2 \cdot I({\rm kA})/R({\rm cm})$, and its direction is azimuthal (the field
lines are toroidal, encircling the inner conductor).  The inner
conductor shape and current were optimized by using GEANT~\cite{GEANT}
to maximize the $\nu_\mu$ flux between 0.5-1~GeV at the detector while
minimizing flux above 1~GeV.  The horn was optimized to run at 170~kA
for $10^8$ pulses with $<3$\% fatigue failure probability.

Despite focusing, a highly divergent hadron beam exits the horn and
enters the decay pipe. This beam consists mainly of unscattered and
scattered primary protons and mesons. The decay pipe is 50~m long and
six feet in diameter; most of the kaons and about a quarter of the
pions decay before reaching its end. At the end of the decay pipe,
50~m from the target, is a beam absorber which stops all the hadrons
and low-energy muons. Located 25~m from the target is an intermediate
absorber which can be lowered into the beam.  This design feature was
introduced to provide a systematic check on muon-decay $\nu_e$
background.

\begin{figure}[t]
\includegraphics[width=3in]{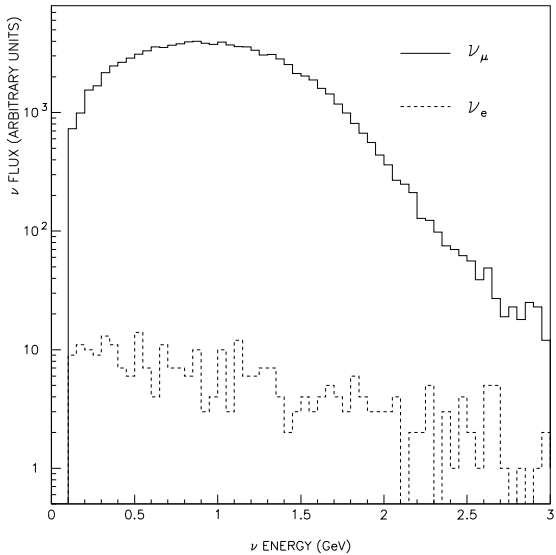}
\caption{The $\nu_{\mu}$ flux (solid) at the MiniBooNE detector
compared to the $\nu_e$ background (dashed).}
\label{fig:fluxnue}
\end{figure}

The neutrino flux which results from this design was simulated using
GEANT with the standard FLUKA hadron interaction package. Efforts to
simulate the flux using the MARS~\cite{mars} simulation package are
underway. All beamline elements, including the horn, shielding, and
absorbers, were simulated. The $\nu_\mu$ flux at 500~m (in the absence
of neutrino oscillations) for a detector on the $z$-axis with 6~m
radius is shown as the solid histogram in Fig.~\ref{fig:fluxnue}.
Using a Gaussian fit, the peak of the spectrum is at
0.94~GeV. Forty-eight percent of the spectrum is in the optimal range,
between 0.3 and 1.0 GeV. The intrinsic $\nu_e$ background is shown
as well; it results primarily from $K^+$, $K^0_L$, and $\mu^+$ decays.
 
\section{DETECTOR DETAILS}

The MiniBooNE detector is a 40-foot spherical tank filled with 800~tons
of clear mineral oil and instrumented with 1520 8-inch PMTs. An optical
barrier mounted 35~cm from the tank wall separates the inner fiducial
region from an outer oil region which is used as a veto. The inner
volume is lined with 1280 PMTs mounted directly on the optical barrier.
The remaining 240 PMTs are mounted in the veto region. Most of the PMTs
were acquired from LSND. 

The detector records the hit arrival time and total charge for each
PMT with $\geq 1$~photoelectron in each 100~ns clock cycle. From the
PMT information, \v{C}erenkov rings and delayed scintillation light
are reconstructed. The oil was not doped with scintillator; the
natural scintillation properties of mineral oil are near optimal for
the experiment.  

The \v{C}erenkov ring allows the track direction and
location to be calculated, as well as providing the primary means of
particle identification. Electrons are identified by their characteristic
ring shape, a result of the electromagnetic shower profile. Muons
have a ring shape characteristic of a penetrating track with little
scattering: the outer edge of the ring is well-defined and the
ring is significantly filled in due to the extended track. Muons are 
further identified by their decay electron, except for the 8\% of $\mu^-$
which are captured by nuclei. Finally,
a source of potential background to the $\nu_e$ signal is neutral-current
production of nucleon resonances which decay to $\pi^0$. The $\pi^0 \rightarrow
\gamma\gamma$ decay produces two electromagnetic showers, each one of which
appears very similar to an electron ring. In asymmetric decays, one ring
may not be reconstructed. The recoil nucleon from the resonance decay,
while usually below \v{C}erenkov threshold, produces additional scintillation 
light which can help distinguish $\pi^0$ events from $\nu_e$. In the end,
the $\pi^0$ misidentification background is expected to be comparable
to or below an anticipated oscillation signal. Muon misidentification levels
should be lower yet.

\section{STATUS AND CONCLUSION}

\begin{figure}[t]
\includegraphics[width=3in]{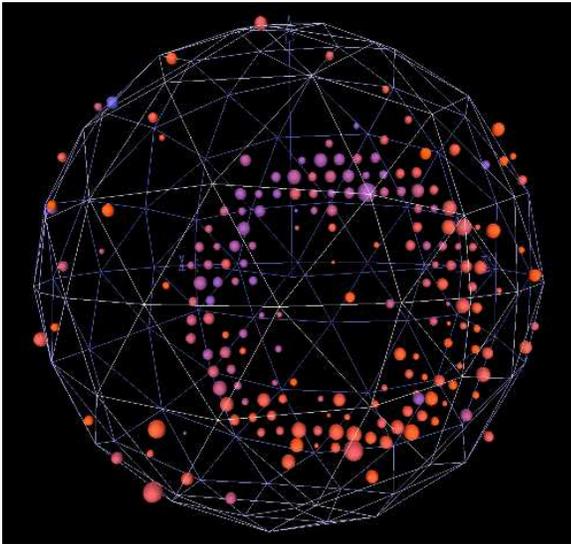}
\caption{Neutrino event from first running period. The colors indicate
hit timing; the size of the colored blobs indicates the total charge
on each PMT.}
\label{evdisp}
\end{figure}

MiniBooNE began taking neutrino physics data in August 2002. As of early
November, several thousand neutrino events have been recorded. A blind 
analysis is being performed, removing $\nu_e$ candidates from the
samples which are open to study. An early $\nu_\mu$ candidate event
display is shown in Fig.~\ref{evdisp}.

The beam and detector were commissioned quickly, and within a few
weeks of the first neutrino data protons were being delivered reliably
at a rate of $10^{16}$ per hour. This is, however, a factor of eight
below the nominal intensity. The Booster's ability to deliver beam is
at present limited by radiation losses in the accelerator and
extraction system.  Work is ongoing to improve this rate, and with
luck the Booster will be delivering nearly the full requested
intensity to MiniBooNE by early in the next calendar year. This
will allow MiniBooNE to publish results addressing LSND in two years.

\end{document}